\documentclass[12pt,preprint]{aastex}

\usepackage{amsmath}

\newcommand{\about}{$\sim\!\!$~}

\newcommand{\be}{\begin{displaymath}}
\newcommand{\ee}{\end{displaymath}}

\def\lsim{\hbox{\rlap{\raise 0.425ex\hbox{$<$}}\lower 0.65ex\hbox{$\sim$}}}
\def\gsim{\hbox{\rlap{\raise 0.425ex\hbox{$>$}}\lower 0.65ex\hbox{$\sim$}}}
\def\arcmin{\hbox{$^\prime$}}

\def\arcdeg{\mbox{$^\circ$}}

\newcommand{\halpha}{H$\alpha$}

\newcommand{\kms}{km~s$^{-1}$}

\graphicspath{{/Users/jsilv/sn2012cg/plots/}}

\shorttitle{The Very Young Type Ia SN 2012cg}
\shortauthors{Silverman, et al.}

\begin{document}

\title{The Very Young Type Ia Supernova 2012cg: Discovery and
  Early-Time Follow-Up Observations}

\author{Jeffrey M. Silverman\altaffilmark{1,2}, Mohan Ganeshalingam\altaffilmark{1}, S. Bradley Cenko\altaffilmark{1}, Alexei V. Filippenko\altaffilmark{1}, Weidong Li\altaffilmark{1,3}, Aaron J. Barth\altaffilmark{4}, Daniel J. Carson\altaffilmark{4}, Michael Childress\altaffilmark{5}, Kelsey I. Clubb\altaffilmark{1}, Antonino Cucchiara\altaffilmark{6}, Melissa L. Graham\altaffilmark{7,8}, 
G. H. Marion\altaffilmark{9}, My L. Nguyen\altaffilmark{4}, Liuyi Pei\altaffilmark{4}, Brad E. Tucker\altaffilmark{5}, Jozsef Vinko\altaffilmark{10,11}, J. Craig Wheeler\altaffilmark{11}, and Gabor Worseck\altaffilmark{5}}


\altaffiltext{1}{Department of Astronomy, University of California, Berkeley, CA 94720-3411, USA.}
\altaffiltext{2}{email: jsilverman@astro.berkeley.edu.}
\altaffiltext{3}{Deceased 12 December 2011.}
\altaffiltext{4}{Department of Physics and Astronomy, University of California, Irvine, CA 92697-4575, USA.}
\altaffiltext{5}{The Research School of Astronomy and Astrophysics, The Australian National University, Mount Stromlo and Siding Spring Observatories, Weston Creek, ACT 2611, Australia.}
\altaffiltext{6}{Department of Astronomy and Astrophysics, UCO/Lick Observatory, University of California, Santa Cruz, CA 95064, USA.}
\altaffiltext{7}{Las Cumbres Observatory Global Telescope Network, 6740 Cortona Dr., Goleta, CA 93117, USA.}
\altaffiltext{8}{Department of Physics, Broida Hall, University of California, Santa Barbara, CA 93106, USA.}
\altaffiltext{9}{Harvard-Smithsonian Center for Astrophysics, 60 Garden St., Cambridge, MA 02138, USA.}
\altaffiltext{10}{Department of Optics and Quantum Electronics, University of Szeged, D\'{o}m t\'{e}r 9, 6720 Szeged, Hungary.}
\altaffiltext{11}{Department of Astronomy, University of Texas, Austin, TX 78712-0259, USA.}

\begin{abstract}
On 2012~May~17.2 UT, only $1.5 \pm 0.2$~d after explosion, we
discovered SN~2012cg, a Type~Ia supernova (SN~Ia) in \object{NGC 4424}
($d \approx 15$~Mpc). As a result of the newly modified 
strategy employed by the Lick Observatory SN Search, a sequence of
filtered images was obtained starting 161~s after discovery. Utilizing
recent models describing the interaction of SN ejecta with a companion
star, we rule out a \about1~M$_\sun$ companion for half of all
viewing angles and a red-giant companion for nearly all
orientations. SN~2012cg
reached a $B$-band maximum of $12.09 \pm 0.02$~mag on 2012~June~2.0
and took \about17.3~d from explosion to reach 
this, typical for SNe~Ia. Our pre-maximum-brightness photometry shows
a narrower-than-average $B$-band light curve for SN~2012cg, though
slightly overluminous at maximum brightness and with
normal color evolution (including some of the earliest
SN~Ia filtered photometry ever obtained). Spectral fits to
SN~2012cg reveal ions typically found in SNe~Ia 
at early times, with expansion velocities $\ga$ 14,000~\kms\ at 2.5~d
past explosion. Absorption from \ion{C}{2} is 
detected early, as well as high-velocity components
of both \ion{Si}{2} $\lambda$6355 and \ion{Ca}{2}. Our last spectrum
(13.5~d past explosion) resembles that of the somewhat peculiar
SN~Ia~1999aa. This suggests that SN~2012cg will
have a slower-than-average declining light curve, which may be 
surprising given the faster-than-average rising light curve. 
\end{abstract}




\section{Introduction}\label{s:intro}

Type~Ia supernovae (SNe~Ia) provided the first clear indication that
the expansion of the Universe is accelerating 
\citep{Riess98:lambda,Perlmutter99} and have been used as precise
distance indicators to accurately measure cosmological parameters 
\citep[e.g.,][]{Conley11,Suzuki12}. Broadly speaking,
SNe~Ia are the result of the thermonuclear explosion of C/O white
dwarfs (WDs); however, the specifics of the progenitor systems and 
explosion mechanisms are still unclear \citep[see][for further
information]{Howell11}.


While there is much value in large statistical studies of SNe~Ia, an
in-depth investigation of an individual object can be extremely
enlightening as well. 
Two recent examples of this are SN~2009ig and SN~2011fe
(PTF11kly). Both are very nearby, normal SNe~Ia that were
discovered 
extremely young \citep{09ig:disc,11fe:disc} and have been intensely
studied by many groups and at all wavelengths. These rich datasets are
excellent test cases for simulations of SNe~Ia and also place strong
observational constraints on many theoretical models 
\citep[e.g.,][]{Foley12,Bloom12,Ropke12,Horesh12,Parrent12}. 

Here we add to this list a nearby SN~Ia
that was found 1.5~d after explosion: SN~2012cg. In
\S\ref{s:discovery} we describe our discovery and 
present our autonomously triggered follow-up 
observations,
initiated mere minutes after discovery. We present our post-discovery
observations and describe our data reduction in
\S\ref{s:observations}, and our analysis of the pre-maximum-brightness
data is discussed in \S\ref{s:analysis}. We summarize our
conclusions in \S\ref{s:conclusions}.


\section{Discovery and Autonomous Follow-Up Observations}\label{s:discovery}

As part of the Lick Observatory Supernova Search
(LOSS; \citealt{Filippenko01}), we have imaged the field of \object{NGC
  4424} with the 0.76~m Katzman Automatic Imaging Telescope
(KAIT; \citealt{Richmond93}) more than 200 times over the last 14 years.
In an 18~s unfiltered KAIT image beginning at 05:21:26 on 2012~May~17
(UT dates are used throughout), we identified a new
transient source with J2000.0 coordinates $\alpha = 12^{\mathrm{h}}
27^{\mathrm{m}} 12\farcs83$, $\delta = +09^{\circ} 25\arcmin
13\farcs1$ \citep[with an uncertainty of 200~mas in each
coordinate;][]{12cg:disc}. 
We found that SN~2012cg had $R = 16.92 \pm 0.05$~mag at the
time of the KAIT discovery (statistical uncertainty only). The next
day the object was spectroscopically classified as an extremely young
SN~Ia \citep{12cg:disc}. 

SN~2012cg is located $17\farcs3$ east and $1\farcs5$ south of
the center of the the peculiar SBa galaxy \object{NGC 4424}, at a
distance of $15.2 \pm 1.9$~Mpc \citep[from the Tully-Fisher
relation,][]{Cortes08}; thus,  SN~2012cg lies a projected distance of
\about1.4~kpc from the host-galaxy center. This places the SN just
outside a central concentration of CO and \halpha\ emission, but among
many blue stars which may indicate recent (but not ongoing) star
formation in the vicinity of SN~2012cg \citep{Cortes06}.

Beginning in early 2011, we modified the LOSS search strategy with the
principal objective of promptly identifying very young (i.e., hours to
days after explosion) SNe in nearby galaxies. In addition to
decreasing the number of galaxies monitored regularly by KAIT (to
increase our observing cadence), we also implemented new software
tools to identify SN candidates in nearly real time. Consequently,
KAIT autonomously obtained a sequence of $U$, $B$, $V$, and 
unfiltered (roughly $R$) images of the transient beginning only 161~s
after the completion of the discovery image. These are some of the
earliest filtered photometry data ever obtained of a SN. Subsequent
filtered (\protect\hbox{$BV\!RI$}) photometry was manually inserted
into the KAIT schedule (see \S\ref{ss:phot}).


\section{Post-Discovery Observations and Data Reduction}\label{s:observations}

\subsection{Photometry}\label{ss:phot}

KAIT continued to monitor SN~2012cg, and the filtered
\protect\hbox{$BV\!RI$} follow-up photometry through \about2.5 weeks
after discovery is analyzed herein. These data were reduced using our
image-reduction pipeline \citep{Ganeshalingam10:phot_paper}. Lacking
host-galaxy templates, point-spread function (PSF) fitting photometry 
was performed at the position of the SN and seven local comparison
stars using the DAOPHOT package in IRAF\footnote{IRAF: The Image
  Reduction 
  and Analysis Facility is distributed by the National Optical Astronomy
  Observatory, which is operated by the Association of Universities
  for Research in Astronomy (AURA) under cooperative agreement with
  the National Science Foundation (NSF).}. Instrumental magnitudes
were transformed to \citet{Landolt92} system magnitudes using
color terms 
derived from multiple photometric nights. The photometric zero point
for each image was determined using the measured magnitudes of the
local standards from Sloan Digital Sky Survey transformed into the
Landolt standard system \citep{Jordi06}.

\subsection{Spectroscopy}\label{ss:spec}

Nearly nightly spectroscopic follow-up observations of SN~2012cg were
also initiated after discovery. Low-resolution optical spectra were
obtained mainly using the Kast double spectrograph on the Shane 3~m
telescope at Lick Observatory \citep{Miller93}, though data were also
acquired with the Marcario Low-Resolution Spectrograph
\citep[LRS;][]{Hill98} on the 9.2~m Hobby-Eberly Telescope (HET) at
McDonald Observatory and the Wide Field Spectrograph (WiFeS, an
integral field unit) on the Australian National University 2.3~m
telescope at Siding Springs Observatory
\citep{Dopita07}. Table~\ref{t:spec} summarizes the spectral
data of SN~2012cg presented here.

\begin{table*}
\begin{center}
\caption{Journal of Spectroscopic Observations of SN~2012cg}\label{t:spec}
\begin{tabular}{lcllrr}
\hline
\hline
UT Date &  Epoch$^\textrm{a}$  &  Instrument & Range (\AA)  & Resolution (\AA)$^\textrm{b}$ &  Exposure (s) \\
\hline
2012~May~18.234 & \phantom{1}2.5 & Kast & 3434--8198 & 4.3/5.7 & 2$\times$1500 \\
2012~May~19.231 & \phantom{1}3.5 & Kast & 3440--8200 & 4.3/5.6 & 2$\times$1200 \\
2012~May~20.226 & \phantom{1}4.5 & LRS &  4200--10200 & 13.5 &  1200 \\
2012~May~20.191 & \phantom{1}4.5 & Kast & 3482--5550 & 4.4 &  1800 \\
2012~May~21.186 & \phantom{1}5.5 & Kast & 3456--10250 & 4.4/10.9 & 2$\times$300 \\
2012~May~22.454 & \phantom{1}6.8 & WIFeS & 3700--9550$^\textrm{c}$ & 0.8/1.3 & 600 \\
2012~May~23.368 & \phantom{1}7.7 & WIFeS & 3700--9550$^\textrm{c}$ & 0.8/1.3 & 600 \\
2012~May~25.200 & \phantom{1}9.5 & LRS & 4200--10200 & 13.5 &  900 \\
2012~May~26.235 & 10.5 & Kast & 3490--8314 & 5.2/5.9 & 300 \\
2012~May~28.198 & 12.5 & Kast & 3490--8124 & 5.1/5.8 & 300 \\
2012~May~29.180 & 13.5 & Kast & 3452--10232 & 4.0/10.5 & 300 \\
2012~May~29.194 & 13.5 & LRS & 4200--10200 & 13.5 &  300 \\
\hline\hline
\multicolumn{6}{p{6.1in}}{$^\textrm{a}$Days relative to the date of
  explosion, 2012~May~15.7. To 
convert to days relative to $B$-band maximum brightness, subtract the
17.3~d rise time from this number (see \S\ref{ss:lightcurves} for more 
information).} \\
\multicolumn{6}{p{6.1in}}{$^\textrm{b}$Approximate full width at
  half-maximum intensity (FWHM) resolution. If two numbers are listed,
  they represent the blue- and red-side resolutions, respectively.} \\ 
\multicolumn{6}{p{6.1in}}{$^\textrm{c}$There is a gap in the spectral
  coverage at 5574--5630~\AA.} \\
\hline\hline
\end{tabular}
\end{center}
\end{table*}

All spectra were reduced using standard techniques
\citep[e.g.,][]{Silverman12:BSNIPI}. Routine 
CCD processing and spectrum extraction were completed with IRAF. We
obtained the wavelength scale from low-order polynomial fits to
calibration-lamp spectra. Small wavelength shifts were then applied to
the data after cross-correlating a template sky to the night-sky lines
that were extracted with the SN. Using our own IDL routines, we fit 
spectrophotometric standard-star spectra to the data in order to flux
calibrate our spectra and to remove telluric lines
\citep{Wade88,Matheson00:93j}.


\section{Analysis and Results}\label{s:analysis}

\subsection{Light Curves}\label{ss:lightcurves}

\begin{figure*}
\centering
\includegraphics[width=5in]{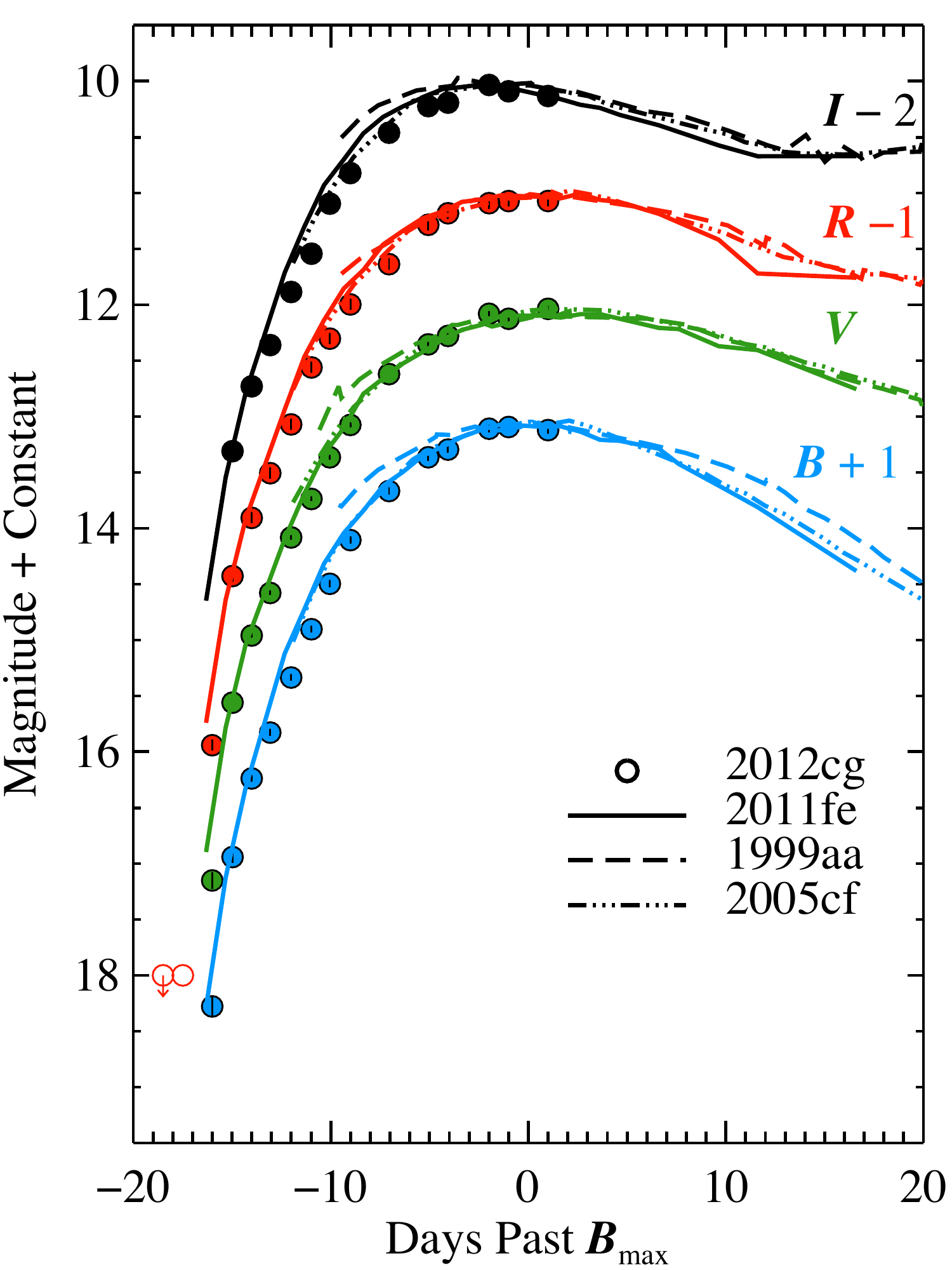}
\caption{\protect\hbox{$BV\!RI$} light curves of SN~2012cg,
  along with comparisons to SNe~Ia 2011fe, 1999aa, and
  2005cf. Comparison light curves have been shifted 
  to have the same peak magnitude and phase as SN~2012cg. Our
  multi-filter data start within minutes
  after our discovery image, \about1.5~d after
  explosion. A nondetection \citep[upper limit,][]{12cg:disc2} and the
  pre-discovery detection \citep[open circle,][]{Lipunov12} are
  also shown.}\label{f:lcs} 
\end{figure*}

We present our \protect\hbox{$BV\!RI$} light curves of SN~2012cg in
Figure~\ref{f:lcs}, compared with those of SN~2011fe (unpublished KAIT
data), 
SN~1999aa \citep{Jha06}, and SN~2005cf \citep{Wang09:05cf}. From a
low-order polynomial fit, we find that SN~2012cg reached a peak
$B$-band magnitude of $12.09 \pm 0.02$ on 2012~June~$2.0 \pm 0.75$.

Assuming $E(B-V)_\textrm{host} = 0.18$~mag (see below) and $d =
15.2$~Mpc \citep{Cortes08}, this implies $M_B = -19.73 \pm 0.30$~mag,
which is \about0.5~mag brighter than a ``normal'' SN~Ia and
\about0.25~mag brighter than the somewhat peculiar SN~1999aa
\citep{Ganeshalingam10:phot_paper}. However, based on our data through
maximum brightness, SN~2012cg appears to have a narrower $B$-band
light curve than the ``normal'' Type~Ia SNe~2011fe and 2005cf and the
slightly overluminous SN~1999aa (even though it resembles this object
spectroscopically; see \S\ref{sss:classification}).

To determine the date of explosion, we model SN~2012cg as an expanding
fireball whose luminosity increases quadratically with time
\citep{Riess99:risetime,Nugent11}. Restricting ourselves to the 5
earliest $B$-band points ($-15.8 < t < -11.8$~d relative to
$B$-band maximum brightness), we find that SN~2012cg was discovered 
$1.5 \pm 0.2$~d after explosion (which occurred on
2012~May~15.7). This is consistent with nondetections on
May~12.3 ($R > 19.8$~mag, KAIT) and May~14.9 \citep[$R >
19.0$~mag,][]{12cg:disc2}, as well as with the faint ``pre-discovery
detection'' on May~15.8 \citep[$R \approx 19.0$~mag;][]{Lipunov12}.
We measure a rise from explosion to maximum light in the $B$ band of 
\about17.3~d, which is within the range of rise times of typical
SNe~Ia \citep[e.g.,][]{Ganeshalingam11}. We note, however, that our
earliest photometry epochs may include a contribution from the host
galaxy due to a lack of galaxy templates. A more accurate measurement
will be made after galaxy templates are obtained and will likely push 
our earliest photometric data slightly fainter than presented here.

Shock heating of SN~Ia ejecta via interaction with a companion is
predicted \citep{Kasen10} to produce thermal emission whose
temperature and luminosity depend both on the binary separation ($a$)
and the observer angle ($\theta$). 
In the case of SN~2012cg, the
early-time ($t < 2$~d) KAIT observations limit any emission
from companion interaction with the SN ejecta to be $\nu L_{\nu} < 
10^{41}$~erg~s$^{-1}$ at optical frequencies.  Using the 
analytic models from \citet{Kasen10}, our limits can rule out
a \about1~M$_\sun$ ($a \approx 2$\,R$_\sun$) main-sequence
companion for  
orientations with $\theta \gtrsim 90^{\circ}$ (i.e., the companion is
perpendicular to our line of sight, or slightly farther away than the
exploding WD) and red-giant companions ($a \approx 400$\,R$_\sun$) for 
all orientations except when the companion is directly behind the WD. 
Analytic models of the emission signature of shock breakout
\citep[e.g.,][]{Chevalier92,Rabinak11} predict very similar
behavior to the (on-axis, or $\theta = 0^{\circ}$) interaction models
of \citet{Kasen10}.  From a similar line of reasoning, our early KAIT
limits also require that the exploding star must have an initial
radius $R_{0} \lesssim 2\, {\rm R}_{\odot}$. 

\begin{figure*}
\centering
\includegraphics[width=5in]{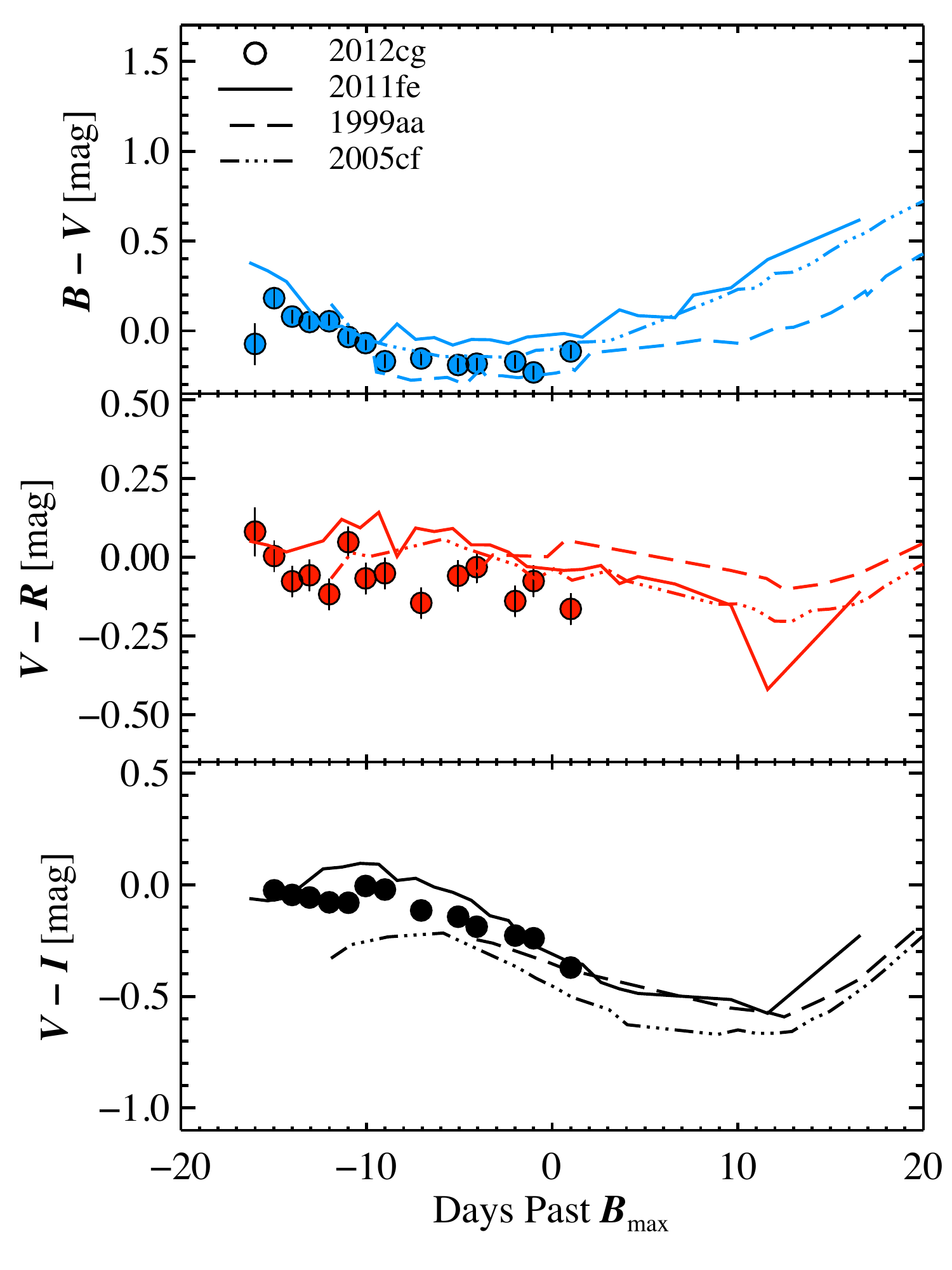}
\caption{Color curves of SN~2012cg, corrected for $E(B-V)_\textrm{MW}
  = 0.019$~mag and $E(B-V) _\textrm{host} = 0.18$~mag. Shown for
  comparison are 
  SNe~2011fe, 1999aa, and 2005cf, all corrected for extinction using
  the reddening values provided in their respective
  references.}\label{f:color}
\end{figure*}

In Figure~\ref{f:color} we plot the evolution of the color of
SN~2012cg with time, along
with some comparison objects. All objects have been corrected for
Galactic extinction using the dust maps of \cite{Schlegel98} and
host-galaxy extinction values from their respective references. 
No host-galaxy extinction is adopted for SN~2011fe. 
In order to match our color curves to those of other SNe~Ia, we
require $E(B-V)_\textrm{host} \approx 0.18$~mag. 
In general, SN~2012cg appears to follow the color evolution of the
``normal'' Type Ia SNe~2011fe and 2005cf, as well as the somewhat
peculiar SN~1999aa. We also note that these are some of the
earliest color data for any SN~Ia.

\subsection{Spectra}\label{ss:spectra}

\begin{figure*}
\centering
\includegraphics[width=5.5in]{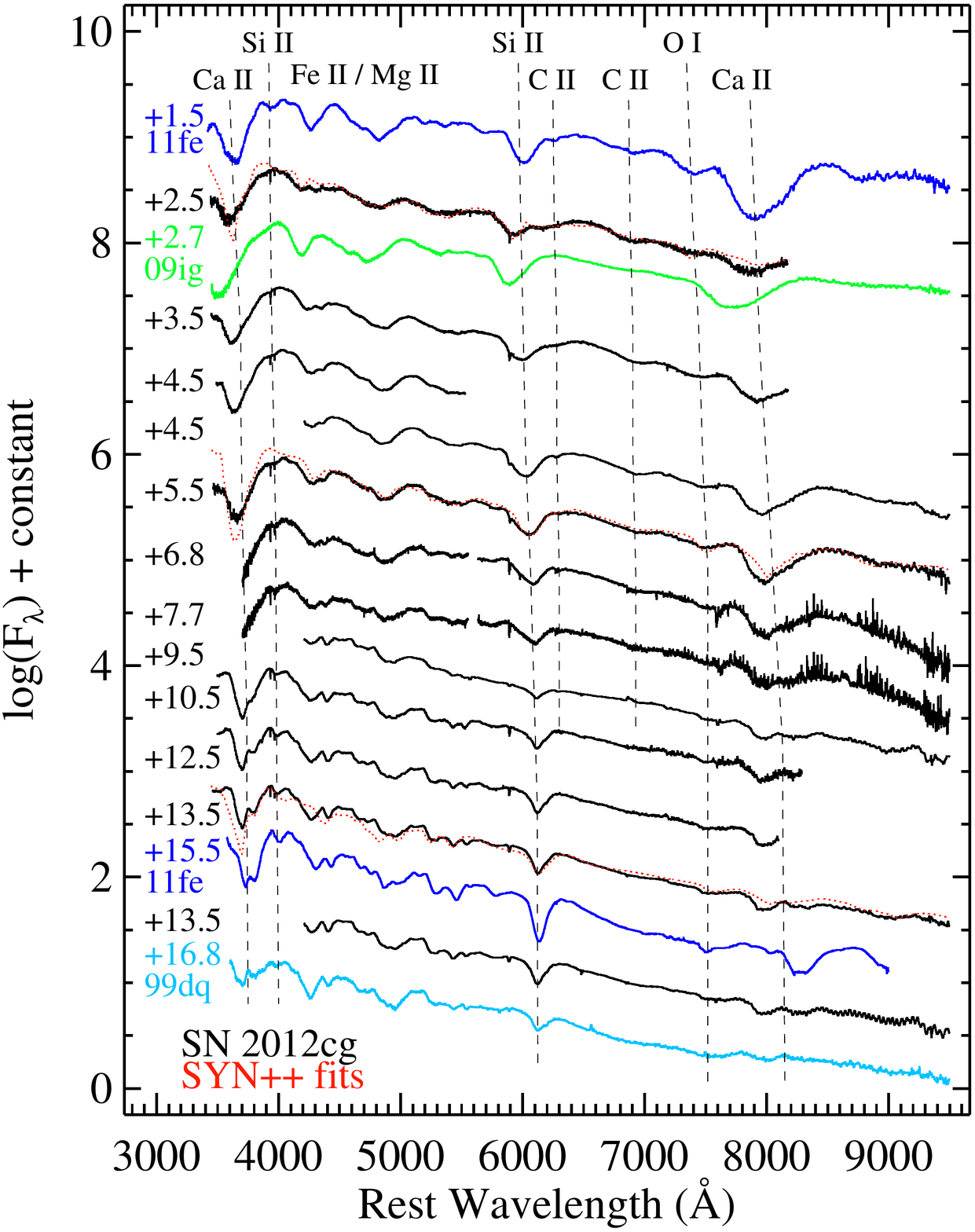}
\caption{Spectra of SN~2012cg ({\it black}) and some {\tt SYN++} fits ({\it
    red}), along with comparisons to other young SNe~Ia: SNe~2011fe
  \citep[{\it dark blue},][]{Parrent12}, 2009ig \citep[{\it
    green},][]{Foley12}, and 1999dq \citep[{\it light
    blue},][]{Silverman12:BSNIPI}. Each spectrum is labeled with its
  age relative to explosion. 
  The data have all been deredshifted and
  dereddened. 
  Major spectral features are identified. 
  The noise at the red end in the \about7 and \about8~d spectra is due
  to incomplete removal of night-sky emission lines.}\label{f:spec} 
\end{figure*}

Our spectra of SN~2012cg from the first
\about2 weeks are shown in Figure~\ref{f:spec}. All of
the spectra exhibit narrow \ion{Na}{1}~D and \ion{Ca}{2}~H\&K absorption
from the host galaxy, though we usually do not
resolve the former. The median redshift as determined from both of
these features is $z = 0.00152 \pm 0.00024$, consistent with the
published redshift of \object{NGC 4424} \citep[$z =
0.00146$;][]{Kent08}. 

The equivalent width (EW) of \ion{Na}{1}~D absorption is often
converted into reddening, but there is some controversy over the
exact relationship \citep[][and references therein]{Poznanski11}. No
\ion{Na}{1}~D absorption is 
detected at $z=0$; thus, we use $E(B-V)_\textrm{MW} = 0.019$~mag, as 
given by \citet{Schlegel98}. The median EW of \ion{Na}{1}~D from the 
host galaxy is measured to be \about0.7~\AA, which yields a range of
possible reddening values (0.00--0.60~mag) with a nominal value of
0.22~mag \citep[with the majority of the uncertainty coming from the
scatter in the relationship presented in][]{Poznanski11}. This is
consistent with $E(B-V)_\textrm{host} \approx 0.18$~mag, the value
derived from the photometry in \S\ref{ss:lightcurves}. 

\subsubsection{SYN++}\label{sss:syn}

To help identify the species present in our spectra of SN~2012cg, we
used the spectrum-synthesis code {\tt SYN++}
\citep{Thomas11:synapps}. {\tt SYN++} is derived from {\tt SYNOW}
\citep{Synow} and is a parameterized resonance-scattering code that
allows for the adjustment of chemical composition, optical depths,
temperatures, and velocities. Three examples of our {\tt SYN++} fits
are shown in Figure~\ref{f:spec}.

Our first spectrum of
SN~2012cg (2.5~d after explosion) consists of absorption features from
ions usually seen in SNe~Ia (\ion{Ca}{2}, \ion{Si}{2},
\ion{Fe}{2}, \ion{Mg}{2}, \ion{S}{2}, and \ion{O}{1}), as well as
\ion{C}{2} which is found in over one-fourth of all SNe~Ia
\citep[e.g.,][]{Silverman12:carbon}. All of these species 
have expansion velocities $\ga$ 14,000~\kms, similar to what
was found in the earliest spectra of SN~2011fe \citep{Parrent12} and
SN~2009ig \citep{Foley12}. 
The \ion{C}{2} features $\lambda$6580 and $\lambda$7234 are detected
in our first spectrum, become more distinct but weaken
with time, and by 10.5~d past explosion the only sign of C is a
flattening of the red wing of the \ion{Si}{2} $\lambda$6355 feature
(presumably due to \ion{C}{2} $\lambda$6580). Just 2~d later no 
\ion{C}{2} is required in our {\tt SYN++} fits. 

In addition to the usual photospheric \ion{Si}{2} absorption,
SN~2012cg exhibits high-velocity (HV) \ion{Si}{2} $\lambda$6355 in its
earliest spectra. This HV absorption appears to be detached from the
rest of the photosphere, with our {\tt SYN++} fits showing a gap of
\about4000~\kms\ separating the normal and HV \ion{Si}{2} $\lambda$6355
features. The HV feature dominates the \ion{Si}{2} $\lambda$6355
profile in our first spectrum \citep[also seen by][]{12cg:disc3}, but
only 1~d later the normal and HV 
features are roughly equal in strength. By \about7~d 
past explosion, the normal-velocity feature dominates the \ion{Si}{2}
$\lambda$6355 profile and our {\tt SYN++} fits to spectra older than
12.5~d past explosion do not require any HV \ion{Si}{2}. This temporal
evolution of the HV \ion{Si}{2} $\lambda$6355 feature almost exactly
mirrors that of SN~2009ig \citep{Foley12}.

Similarly, \ion{Ca}{2} also displays HV features in our first
spectrum. Both the \ion{Ca}{2} H\&K feature and the \ion{Ca}{2}
near-IR triplet show absorption profiles that consist of a
photospheric component as well as a HV component with a velocity gap
of 4000--8000~\kms. In contrast to the HV \ion{Si}{2}, the HV
\ion{Ca}{2} is required in {\it all} of our {\tt SYN++} fits, through
13.5~d past explosion. HV \ion{Ca}{2} was also seen in spectra of
SN~2011fe younger than \about8~d past explosion \citep{Parrent12}. 


\subsubsection{Individual Line Measurements}\label{sss:lines}

\begin{figure*}
\centering$
\begin{array}{c}
\includegraphics[width=5in]{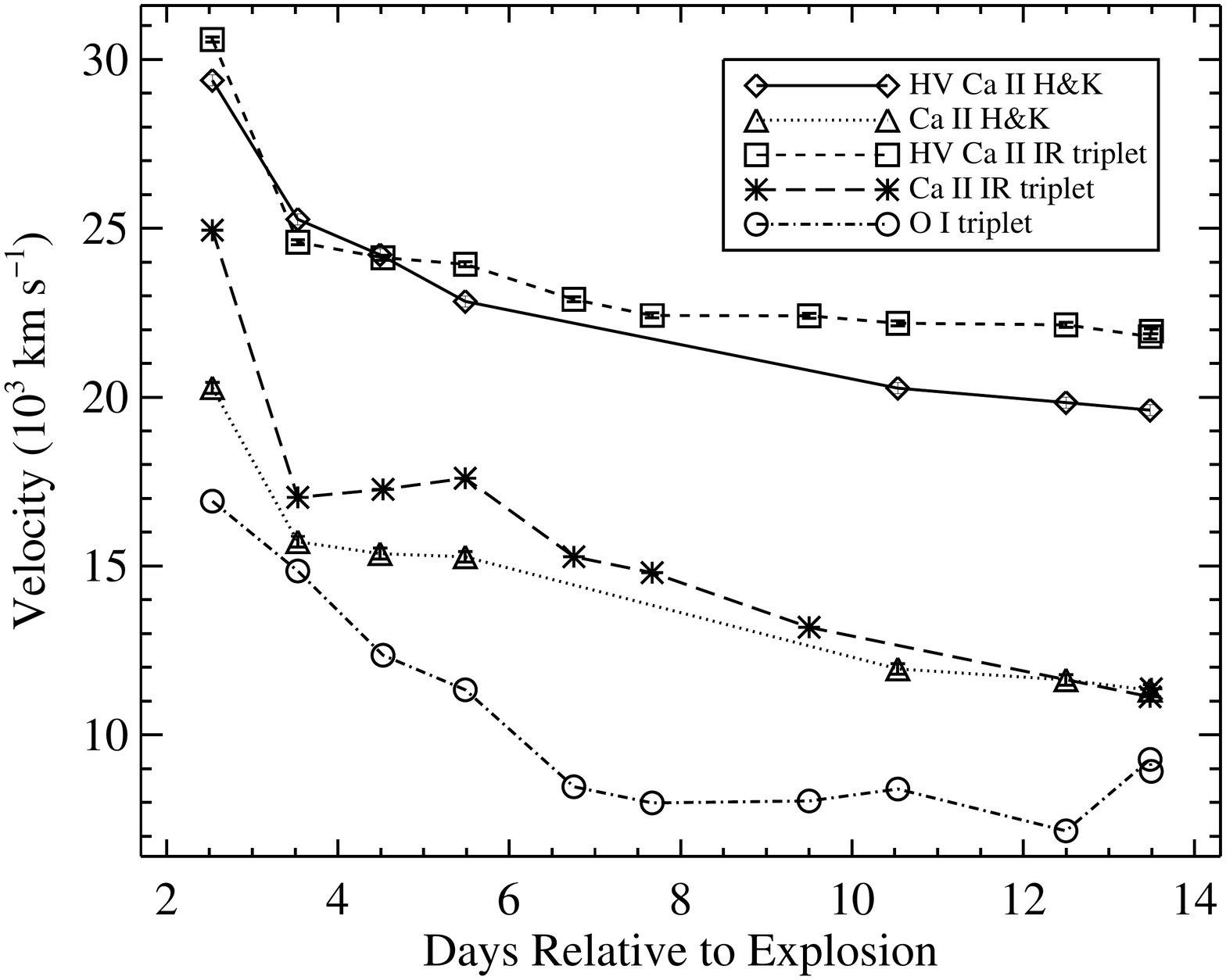} \\
\includegraphics[width=5in]{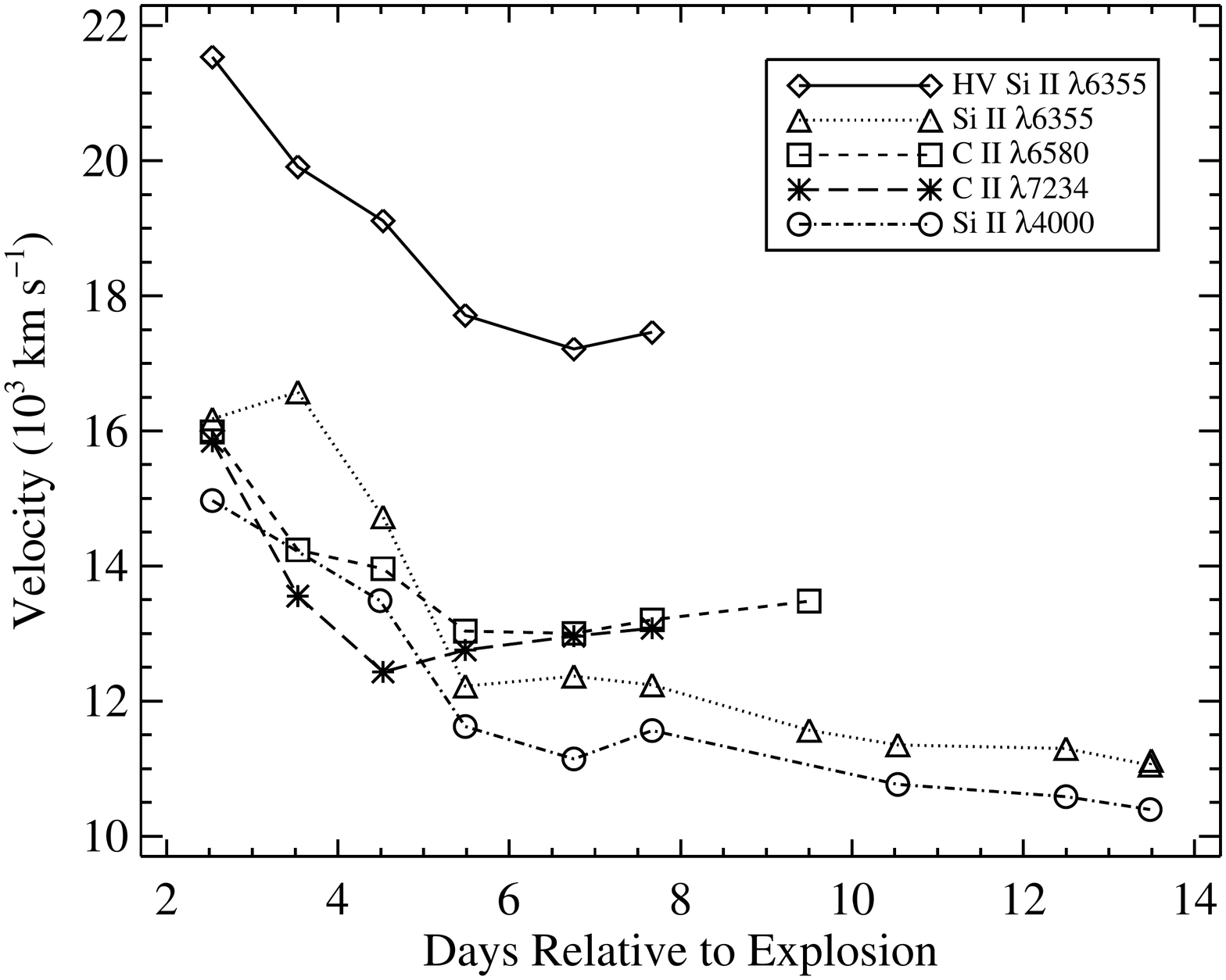}
\end{array}$
\caption{The temporal evolution of the expansion velocity of 
  \ion{Ca}{2} (both HV and photospheric components) and \ion{O}{1}
  features ({\it top}) and of \ion{Si}{2} (both HV and photospheric
  components) and \ion{C}{2} features ({\it bottom}). Uncertainties
  are smaller than the size of the data points.}\label{f:lines}
\end{figure*}

To more precisely study the evolution of our spectra of SN~2012cg, we
directly measured the expansion velocities and EWs of some of the
absorption features; see \citet{Silverman12:BSNIPII} for details.  
All of the
measured velocities are consistent with our {\tt SYN++} fits and are
plotted in Figure~\ref{f:lines}.

As mentioned previously, all of our SN~2012cg spectra show evidence
for HV \ion{Ca}{2}. In our first spectrum, 2.5~d past explosion,
the \ion{Ca}{2} H\&K feature and \ion{Ca}{2} near-IR triplet have 
velocities of \about30,000~\kms, which is even higher than in the
earliest spectra of SN~2011fe \citep{Parrent12}. As seen in the top
panel of Figure~\ref{f:lines}, the two \ion{Ca}{2} features have very
similar velocity evolution in both the HV and photospheric
components. 

Also plotted in the top panel of Figure~\ref{f:lines} is the velocity
evolution of the \ion{O}{1} triplet. Initially, these velocities are
relatively large and similar to the photospheric \ion{O}{1} velocity
seen in the earliest spectra of SN~2011fe. However, that SN also
showed HV \ion{O}{1} earlier than 5~d after explosion which is absent
from SN~2012cg \citep{Parrent12}. 

As discussed in \S\ref{sss:syn} and seen in the bottom panel of
Figure~\ref{f:lines}, SN~2012cg exhibited a HV \ion{Si}{2}
$\lambda$6355 feature in the earliest observations. Our spectrum from
2.5~d past explosion shows the HV \ion{Si}{2} $\lambda$6355 to be
at \about22,000~\kms, which is slightly lower than the HV \ion{Si}{2}
seen in either SN~2009ig \citep{Foley12} or SN~2011fe \citep{Parrent12}. 
If we
associate the steepening of the red edge of the \ion{Ca}{2} H\&K
feature in our earliest spectra 
\citep[which was also seen in SN~2009ig;][]{Foley12}  
with absorption from \ion{Si}{2}
$\lambda$4000, the velocity of this feature is consistent with that
of the \ion{Si}{2} $\lambda$6355 line.

Also in the bottom panel of Figure~\ref{f:lines}, we plot the velocity
evolution of the \ion{C}{2} features $\lambda$6580 and
$\lambda$7234.  
The velocity of \ion{C}{2} $\lambda$7234 is
slightly lower than that of \ion{C}{2} $\lambda$6580 and quite
similar to that of the photospheric component of \ion{Si}{2}
$\lambda$6355, which has been seen in previous work
\citep[e.g.,][]{Silverman12:carbon}. 
The ratio of the
velocity of \ion{C}{2} $\lambda$6580 to \ion{Si}{2} $\lambda$6355
decreases slightly with time, though in each spectrum this ratio is
within 10\% of the median value found for a large sample of SNe~Ia
\citep[e.g., 1.05;][]{Silverman12:carbon}.
We also measure the EW of
both of the \ion{C}{2} lines and find that 
there is some evidence for a factor of 2--3
increase in the EW of \ion{C}{2} $\lambda$6580 from 5.5 to 9.5~d past 
explosion, but this is significantly smaller than the factor of
\about5 increase at similar epochs observed in one object
\citep[SN~1994D,][]{Silverman12:carbon} and predicted from synthetic
spectra \citep{Folatelli12}.

\subsubsection{Classification}\label{sss:classification}

Using the SuperNova IDentification code
\citep[SNID;][]{Blondin07} along with the spectral templates described
by \citet{Silverman12:BSNIPI}, we find that SN~2012cg is
spectroscopically similar to the somewhat peculiar SN~1999aa and
its ilk \citep{Li01:pec,Strolger02,Garavini04}. SNID indicates that
several of the oldest spectra presented here are most similar to
those of SN 1999aa-like objects (at the correct redshift and age). 
\citet{Branch06} utilized the near-maximum brightness EWs of
\ion{Si}{2} $\lambda$6355 and \ion{Si}{2} $\lambda$5972 in order to
classify SN~Ia spectra. Using the oldest spectrum presented in this work, we
find that both of these features are weak and thus
SN~2012cg definitely falls in the Shallow Silicon (SS) class (as do
SN~1999aa and its brethren). 

The oldest spectrum of SN~2012cg presented here (from
13.5~d after explosion) vaguely resembles both SNe~2009ig
(not shown) and 2011fe at similar epochs. However, SN~2012cg has
significantly weaker \ion{Si}{2} absorption and still shows HV
\ion{Ca}{2}. As seen at the bottom of Figure~\ref{f:spec}, a much
better match to this spectrum of SN~2012cg is the SN~1999aa-like
SN~1999dq from \about17~d past explosion
\citep{Silverman12:BSNIPI,Ganeshalingam11}. Thus, our spectral data
indicate that SN~2012cg is a SN~1999aa-like object. This, along with
the fact that SN~2012cg is overluminous at maximum brightness
(\S\ref{ss:lightcurves}), suggests that it will have a
slower-than-average light-curve decline.\footnote{Post-maximum
  photometric data indicate $\Delta m_{15} \approx 0.83$~mag, which
  is indeed slower than average.}


\section{Conclusions}\label{s:conclusions}



In this {\it Letter} we presented optical photometry and spectroscopy
of SN~2012cg, a SN~Ia discovered a mere $1.5 \pm 0.2$~d after
explosion. It is found to be overluminous at maximum brightness and
spectroscopically similar to the somewhat peculiar SN~Ia~1999aa. These
properties imply that SN~2012cg will have a slowly declining light
curve. However, the pre-maximum-brightness photometry 
indicates a faster-than-average rising light curve. Future
observations will be required to definitively
determine the relative normalcy (or peculiarity) of SN~2012cg.

This object will certainly become one of the best-studied SNe~Ia
ever. In under 3~years there have been three very nearby SNe~Ia
discovered within days after explosion (SNe~2009ig, 2011fe, and
2012cg). During that time transient surveys of various sizes
and cadences have continued to grow and evolve, and yet these
extremely young SNe~Ia are still rare finds. Therefore, they will serve
as excellent case studies well into the future and will help us answer
some of the many outstanding questions in the field of SNe~Ia that
plague us today.

\begin{acknowledgments}
We would like to thank J.~Caldwell, J.~R.~Mould, S.~Odewahn,
J.~X.~Prochaska, S.~Rostopchin, M.~Shetrone, and I.~Shivvers for their
assistance with some of the observations.
We are grateful to the staff at Lick Observatory for their support and
are indebted to the benefactors, builders, and partners of the HET 
and LRS.
This work was supported in part by NSF grant AST--1109801 and
Hungarian grant OTKA K76816. A.V.F.'s group at UC Berkeley, and KAIT
and its ongoing operation, have received financial assistance from the
Sylvia \& Jim Katzman Foundation, Gary \& Cynthia Bengier, the Richard
\& Rhoda Goldman Fund, the TABASGO Foundation, and NSF grant AST-0908886.
\end{acknowledgments}


\end{document}